\documentclass{PoS}

\usepackage{latexsym,amssymb,amsfonts,amsmath,slashed}
\newcommand{\f}[2]{\frac{#1}{#2}}
\newcommand{\la}{\langle}

\newcommand{\ra}{\rangle}

\newcommand{\slaD}{{\slashed D}}
\newcommand{\compresslist}{%
\setlength{\itemsep}{1.5pt}%
}

\title{Understanding localisation in QCD through an Ising-Anderson model}

\ShortTitle{Understanding localisation in QCD through an Ising-Anderson model}

\author{\speaker{Matteo Giordano}\thanks{Supported by the Hungarian
    Academy of Sciences under ``Lend\"ulet'' grant
    No. LP2011-011.}\\
  Institute for Nuclear Research of the Hungarian Academy of
  Sciences,\\
  Bem t\'er 18/c, H-4026 Debrecen, Hungary\\
  E-mail: \email{giordano@atomki.mta.hu}}

\author{Tam\'as G.\ Kov\'acs\footnotemark[2]\\
  Institute for Nuclear Research of the Hungarian Academy of
  Sciences,\\
  Bem t\'er 18/c, H-4026 Debrecen, Hungary\\
  E-mail: \email{kgt@atomki.mta.hu}}
\author{Ferenc Pittler\\
  MTA-ELTE Lattice Gauge Theory Research Group,\\ P\'azm\'any
  P. s\'et\'any 1/A, H-1117 Budapest, Hungary\\
  E-mail: \email{pittler@bodri.elte.hu}}

\abstract{Above the QCD chiral crossover temperature, the low-lying
  eigenmodes of the Dirac operator are localised, while moving up in
  the spectrum states become extended. This
  localisation/delocalisation transition has been shown to be a 
  genuine second-order phase transition, in the same universality
  class as that of the 3D Anderson model.   
  The existence of localised modes and the effective dimensional
  reduction can be tentatively explained as a consequence of local
  fluctuations of the Polyakov loop, that provide 3D on-site
  disorder, in analogy to the on-site disorder of the Anderson model.  
  To test the viability of this explanation we study a 3D effective,
  Anderson-like model, with on-site disorder provided by the spins of 
  a spin model, which mimics the Polyakov loop dynamics. Our preliminary
  results show that localised modes are present in the ordered phase,
  thus supporting the proposed mechanism for localisation in QCD.
}

\FullConference{The 32nd International Symposium on Lattice Field
  Theory,\\ 
  23-28 June, 2014\\
  Columbia University New York, NY}

\begin{document}

\section{Introduction}

Low-lying Dirac eigenmodes are of considerable physical interest,
especially in connection with the issue of chiral symmetry breaking. 
In recent years, it has become clear that the localisation properties
of the low-lying modes change across the chiral crossover: while at
low temperatures all the Dirac eigenmodes are delocalised, above the
chiral crossover temperature $T_c$ the low modes become localised on
the scale of the inverse temperature; higher up in the spectrum, above
a critical eigenvalue $\lambda_c(T)$, modes remain delocalised also
above $T_c$~\cite{GGO,GGO2,KGT,KP,KP2,feri,crit}. 

The coexistence of localised and delocalised modes in the same
spectrum is a well known phenomenon in condensed matter physics, and
it is the characteristic feature of the 3D Anderson model for ``dirty''
conductors~\cite{Anderson58,LR,EM}. The Hamiltonian of the Anderson model
consists of the usual tight-binding Hamiltonian plus a random on-site
potential, 
\begin{equation}
  \label{eq:andersonmodel}
  H_{\vec x\, \vec y} = \varepsilon_{\vec x} \delta_{\,\vec x \,\vec y} + \sum_{\mu} 
  (\delta_{\,\vec x+\hat\mu\,\vec y} +
  \delta_{\,\vec x-\hat\mu\,\vec y})\,,
\end{equation}
where $\varepsilon_{\vec x}$ 
provides a diagonal disorder term mimicking the presence of 
impurities in the crystal. The random potential is usually chosen to
be uniformly distributed in an interval $\varepsilon_{\vec x}\in
\left[-\f{W}{2},\f{W}{2}\right]$, with the width $W$ controlling the
amount of disorder in the system. If a magnetic field is applied
to the system, the model is modified to 
\begin{equation}
  \label{eq:uandersonmodel}
  H_{\vec x \,\vec y} = \varepsilon_{\vec x} \delta_{\,\vec x\, \vec y} + \sum_{\mu} 
  (\delta_{\,\vec x+\hat\mu\,\vec y} +
  \delta_{\,\vec x-\hat\mu\,\vec y})
{e^{i\phi_{\vec x \,\vec y}}\,,
\qquad\phi_{\vec y\, \vec x}=-\phi_{\vec x \,\vec y}}\,,
\end{equation}
where $\phi_{\vec x\, \vec y}$ are random phases. This is the 
unitary Anderson model. The name ``unitary'' comes from the symmetry 
class to which the Hamiltonian Eq.~\eqref{eq:uandersonmodel} belongs in
the random matrix theory classification. In the same classification,
the model Eq.~\eqref{eq:andersonmodel} belongs to the orthogonal
class. In both these models, for any nonzero $W$, there
is a critical energy $E_c(W)$, called ``mobility edge'', which separates
localised and delocalised modes: for energies beyond $E_c$, the energy
eigenmodes  at the band edge are localised, while eigenmodes in the
band center are delocalised. The transition from
localised to delocalised modes as one moves along the spectrum is a
true second-order phase transition, known as Anderson transition. 

It has been recently shown that the analogous transition in the Dirac 
spectrum above $T_c$ is also a genuine second-order phase
transition~\cite{crit}. The critical exponent of the correlation
length has been determined, and found to be $\nu_{\rm QCD} = 1.43(6)$,
which is compatible with the one found in the 3D unitary Anderson
model, $\nu_{\rm UAM}=1.43(4)$~\cite{nu_unitary}. While the appearence
of the unitary class is expected, since it is the symmetry class of
QCD in the language of random matrix models, the fact that a phase
transition in a 4D theory seems to belong to the same universality
class as a phase transition in a 3D model needs to be explained. The
dimensionality of the system is not the only difference between 
high-temperature QCD and the Anderson model. While in the Anderson
model disorder is diagonal\footnote{Although in the unitary model
  there is also off-diagonal disorder, this is known to be much less
  effective in inducing localisation~\cite{offdiag}.} and
uncorrelated, in QCD it is off-diagonal, i.e., appearing in the
hopping terms, and correlated. Correlations are short-range, so they
should not be relevant; on the other hand, how off-diagonal disorder
in QCD can produce the same effects as diagonal disorder in the
Anderson model requires an explanation. 

In this contribution we propose a mechanism for localisation in
QCD, which is able to explain the apparent differences between
the two models. The argument consists in a refinement of the proposal
of Ref.~\cite{Bruckmann:2011cc}. To test this mechanism, we introduce
and justify an effective model (``Ising-Anderson model''), which
should produce localisation precisely through this same mechanism. We
also show some preliminary numerical results that support the
viability of our explanation.

\section{Polyakov lines and localisation}

In order to better understand the relation between QCD and the Anderson
model, the key observation is that QCD above $T_c$ is effectively a 3D
model, at least as far as the qualitative features of quark
eigenfunctions are concerned. Indeed, as the size of the temporal
dimension is finite and smaller than the correlation length, 
the time slices are strongly correlated, and so quark eigenfunctions
will look qualitatively the same on all time slices, in particular for
what concerns the localisation properties. 

Let us discuss this point in more detail. It is convenient to work in
the temporal gauge, $U_4(t,\vec x)=1$ for $t=0,\ldots,N_T-2$ and
$\forall\,\vec x$, in which $U_4(N_T-1,\vec x)$ equals the local Polyakov
line $P(\vec x) = \prod_{t=0}^{N_T-1} U_4(t,\vec x)$. A further
time-independent gauge transformation allows to diagonalise each local
Polyakov line: we will refer to this as the diagonal temporal
gauge. In any temporal gauge, covariant time differences are replaced
by ordinary differences; the price to pay is that the antiperiodic
boundary conditions become effective, $\vec x$-dependent boundary
conditions, which involve the local Polyakov line,
\begin{equation}
  \label{eq:eff_bc}
\psi(N_T,{\vec x})=-P({\vec x})\psi(0,{\vec x})\,.  
\end{equation}
Since the time slices are strongly correlated, these effective,
$\vec x$-dependent boundary conditions will affect the behaviour at
the spatial point $\vec x$ for all times $t$. Furthermore, $P({\vec
  x})$ fluctuates in space (and obviously from one configuration to
another). From the point of view of a disordered system, QCD above
$T_c$ therefore contains effectively a diagonal (on-site), 3D source
of disorder. 

To see how the effective boundary conditions affect the quark wave
functions, let us first discuss a simplified setting in which the
Dirac equation can be explicitly solved, generalising the argument of
Ref.~\cite{Bruckmann:2011cc} to $SU(3)$. Consider configurations with
constant temporal links $U_4$ and trivial spatial links $U_j=1$. In
the diagonal temporal gauge, one has $P(x)=P={\rm
  diag}(e^{i\varphi_1},e^{i\varphi_2},e^{i\varphi_3})$ with
$\varphi_3=-\varphi_1-\varphi_2$, so the Dirac operator is diagonal in
colour, and the colour components $\psi_k$ of the quark eigenfunctions
decouple. The eigenfunctions of the Dirac operator are plane waves,
\begin{equation}
  \label{eq:eigen_simple}
  \slaD \psi_k = i\lambda\psi_k\,,\quad
\psi_k(t,\vec x) \propto e^{i\omega_k t + i\vec p \cdot\vec
  x}\,, \quad \lambda = \pm\sqrt{\sin^2\omega_k + \textstyle\sum_{j=1}^3\sin^2
p_j}\,,
\end{equation}
with $\f{L}{2\pi}p_j=0,1,\ldots,L-1$ to fulfill the spatial periodic
boundary conditions, and  
 \begin{equation}
   \label{eq:eff_Mats}
\omega_k = \textstyle\f{1}{N_T}(\pi 
+  \varphi_k \mod 2\pi)=aT(\pi 
+  \varphi_k \mod 2\pi)  \,,
 \end{equation}
to fulfill the effective temporal boundary conditions. We will refer
to $\omega_k$ as the effective Matsubara frequencies. For a given
value of the phase $\varphi_k$, $N_T/2$ (twofold degenerate) different
branches for $\omega_k$ appear, corresponding to increasing values of
$\lambda$. For the lowest branch, 
$\omega_k$ decreases as $\varphi_k$ moves away from zero, and it is
minimal when $e^{i\varphi_k}=-1$. 

Let us now consider the general case. Above $T_c$, in a typical gauge
configuration the Polyakov line $P({\vec x})$ gets ordered along 
$\mathbf{1}={\rm diag}(1,1,1)$ with ``islands'' of  ``wrong'' $P({\vec
  x})\neq \mathbf{1}$. If it were completely ordered, there would be a
sharp gap in the spectrum at $\lambda_g=\omega_k(0)=\pi aT$ (and a
symmetric one at $-\lambda_g$), but fluctuations (both of $P(x)$ and
of the spatial links) smoothen it out. In particular, living on the
``islands'' of ``wrong'' $P({\vec x})$ is ``energetically'' favourable
for the quark eigenfunctions, yielding $|\lambda|<\lambda_g$ as long
as the momentum required to localise the state does not overbalance
the gain. The gap becomes therefore an effective gap $\lambda_c$, 
identified as the ``mobility edge'' separating localised and
delocalised modes. This effective gap is furthermore displaced by the
presence of spatial fluctuations, which most likely have a
delocalising effect on the eigenmodes, thus pushing the effective
gap down. 

In summary, the presence of ``islands'' of ``wrong'' Polyakov lines
provides a localising ``trap'' for eigenmodes. At fixed lattice
spacing, we expect $\lambda_c$ to increase as the temperature is
increased, being ``dragged'' by the effective Matsubara frequency.

\section{Effective 3D model}

The considerations above suggest that it should be possible to
understand the qualitative features of the Dirac spectrum and
eigenfunctions in QCD, in particular concerning spectral statistics  
and localisation properties, by using a genuinely 3D model. To
construct such a model, one has to strip off all the features that are
irrelevant to localisation. The first step is to get rid of the time
direction, reducing the lattice to three dimensions, and
replacing the time covariant derivative in the Dirac operator with a 
diagonal noise term, intended to mimic the effective boundary
conditions. Moreover, it is known that off-diagonal disorder is less
effective than diagonal disorder in producing
localisation~\cite{offdiag}, so we can replace spatial covariant
derivatives with ordinary derivatives. As a 
consequence, colour components decouple: this changes the symmetry
class (in the sense of random matrix models), but should not affect 
the presence of localisation. In conclusion, the main features of
localisation should still be captured if we replace the 4D lattice
with a 3D lattice, and the Dirac operator with the effective
``Hamiltonian'' 
\begin{equation}
\label{eq:H}
    i\slaD_{x\,y} = i\gamma^4 (D_4)_{x\,y} + i\vec\gamma
    \cdot\vec D_{x\,y}
    \longrightarrow
    {H_{\vec x \,\vec y} = \gamma^4 {\cal N}_{\vec x}\delta_{\,\vec x
        \,\vec y}} {+ i\vec\gamma 
      \cdot \vec\partial_{\vec x\,
    \vec y}}\,,
\end{equation}
where ${\cal N}_{\vec x}$ is the diagonal noise, to be specified later.
This Hamiltonian is diagonal in colour, and so has effectively only
spacetime and Dirac indices; for lattices of even spatial size, a spin
diagonalisation allows to get rid of the latter.

At this point we should specify the diagonal noise ${\cal N}_x$
intended to mimic the effective boundary conditions, which has
therefore to satisfy a few requirements:
\begin{enumerate}
\item it should not be uncorrelated, but rather be governed by
  Polyakov-loop-like dynamics. This suggests to base it on some spin
  model in the ordered phase; 
\item as it is the phase of $P({\vec x})$ which enters the effective
  boundary conditions, which is a continuous variable, we have to
  use continuous spins;
\item finally, ${\cal N}_{\vec x}$ should produce an effective gap in
  the spectrum.
\end{enumerate}
For our purposes it is enough to have a continuous spin model which
displays an ordered phase, and the simplest choice is the Ising model
with continuous spin. A possible choice for the noise term is
\begin{equation}
  \label{eq:noise}
  {\cal N}_{\vec x} =\Lambda\f{1+s_{\vec x}}{2}\,, \qquad s_{\vec x}\in[-1,1]\,,
\end{equation}
where $s_{\vec x}$ is the spin variable at point $\vec x$, and
$\Lambda$ is a constant 
determining the strength of the coupling of the fermions to the spins.
Since $\frac{1+s_{\vec x}}{2}$ is 1 for ``aligned'' spins (i.e., for
$s_{\vec x}=1$), this choice provides indeed an effective spectral
gap when the spins are ordered.\footnote{It is understood that we work
  with magnetic field $h=0^+$.} 

Putting everything together, our effective model reads
\begin{equation}
  \label{eq:H2}
  H_{\vec x\,\vec y} = \gamma^4
\Lambda\textstyle\frac{1+s_{\vec x}}{2}\delta_{\,\vec x\,\vec y} + i\vec\gamma
\cdot\vec \partial_{\vec x\,\vec y}\,,
\end{equation}
with $s_{\vec x}$ distributed according to the dynamics of the Ising model,
\begin{equation}
  \label{eq:ising}
  \f{H_{\rm Ising}}{kT} = -\beta_{\rm Ising} \sum_{<\vec x\,\vec y>}
  s_{\vec x} s_{\vec y}\,, \qquad s_{\vec x}\in[-1,1]\,.
\end{equation}
In the ordered phase of the Ising model there is a ``sea'' of $s_{\vec
  x}=1$ spins with ``islands'' of $s_{\vec  x}\neq 1$ spins, so the
underlying configurations have the same features as the Polyakov loop
configurations in QCD. While in QCD we have a single parameter
governing the ordering of the configuration and the size of the
effective gap, in the effective model the ordering of the spin
configuration is governed by $\beta_{\rm Ising}$, while the size of
the gap is mainly determined by the spin-fermion coupling $\Lambda$. 

A curious feature of this model is that it belongs to different
symmetry classes for lattices of even or odd size $L$, namely to the
orthogonal class for $L$ even, and to the symplectic class for $L$
odd. As we have already said above, it is more convenient to work with
even-sized lattices, since one can get rid of the Dirac indices
through a spin diagonalisation. Studying the localisation properties
of this model would provide a test for the viability of the
sea/islands explanation.

\begin{figure}[t]
  \centering
  \includegraphics[width=0.48\textwidth]{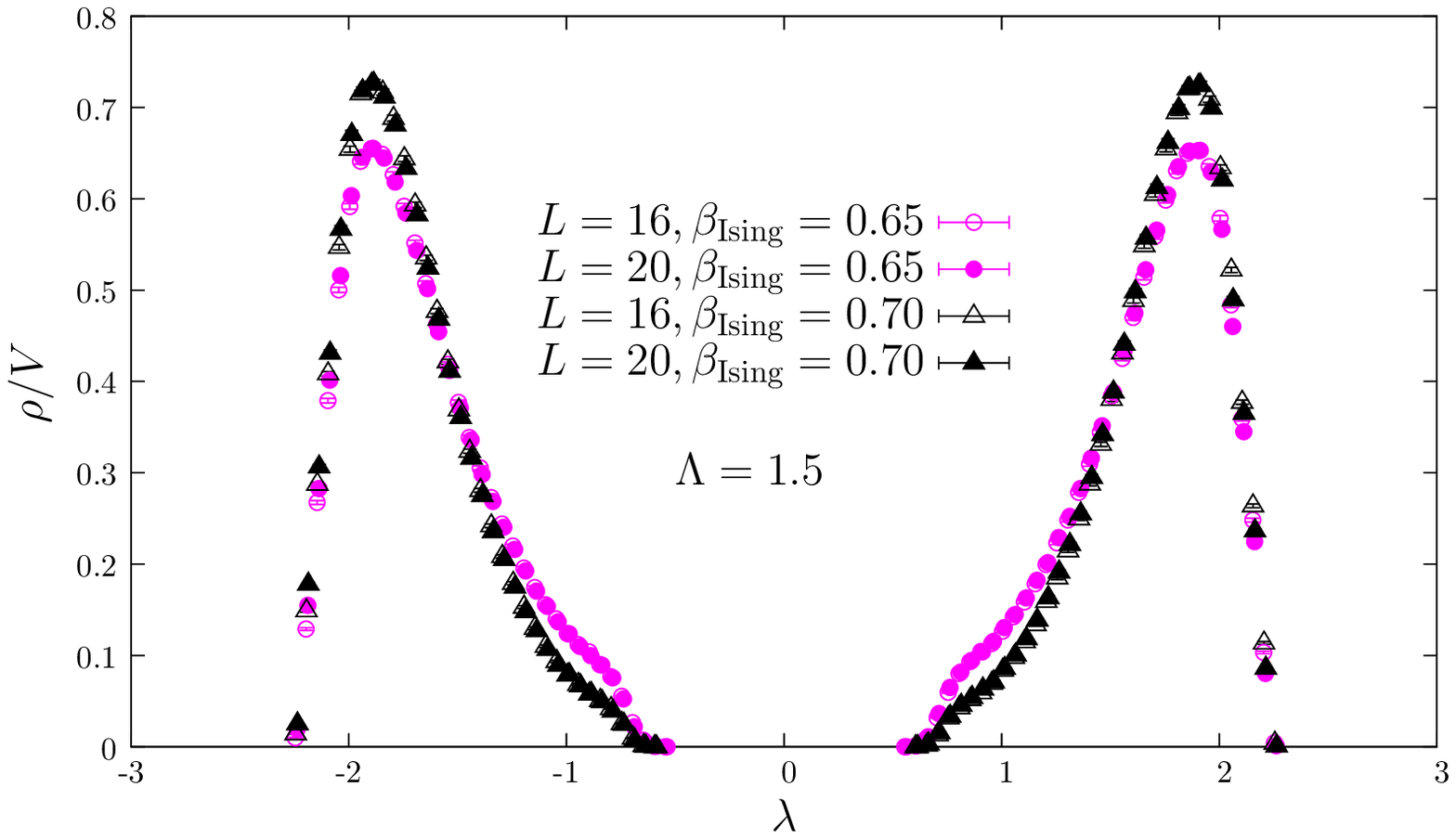}\hfil
  \includegraphics[width=0.48\textwidth]{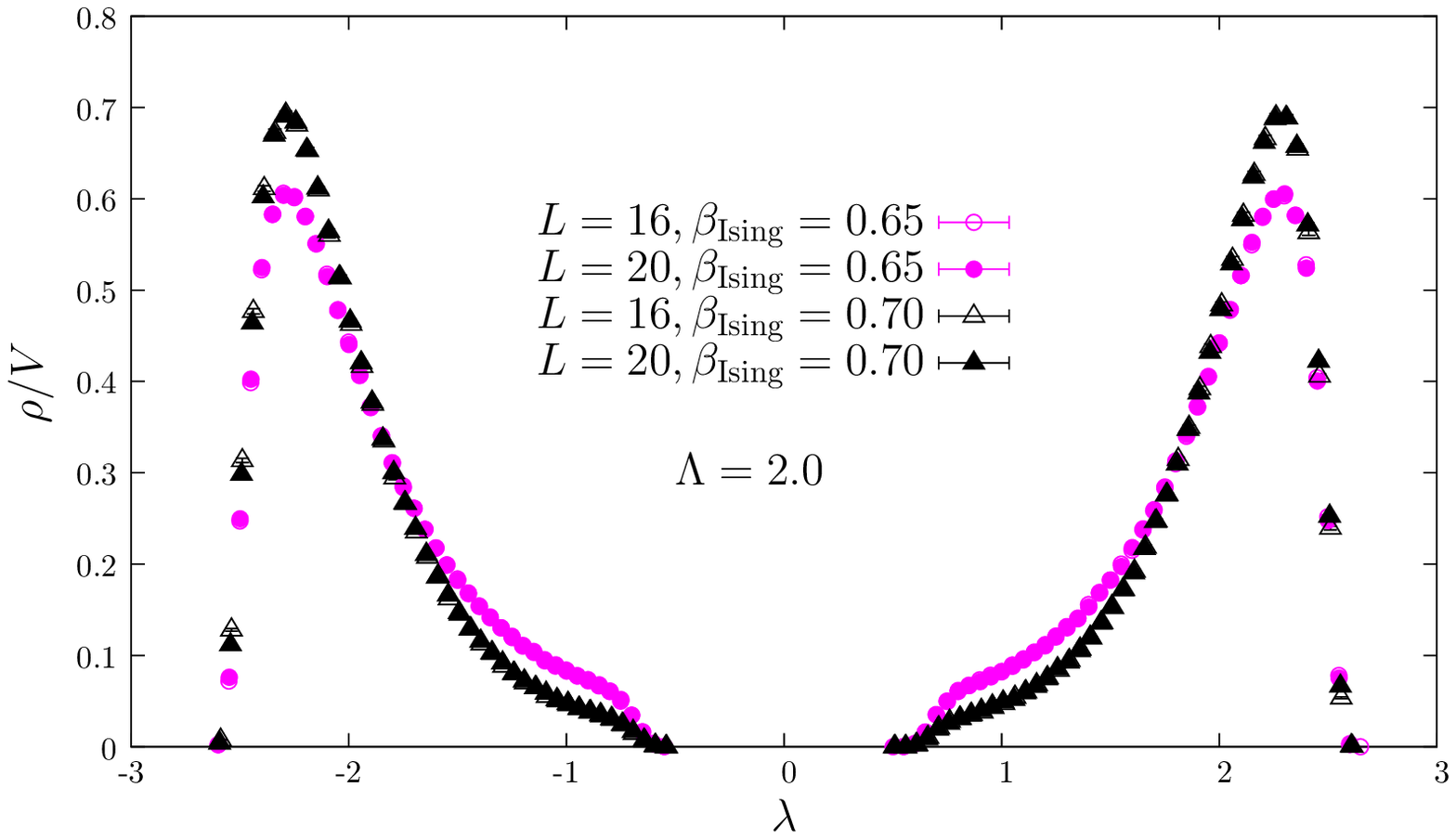}
  \caption{Spectral density of the effective model Eq.~\protect\eqref{eq:H2}.}
  \label{fig:1}
\end{figure}

\section{Numerical results}

We have performed numerical simulations of the effective model
Eq.~\eqref{eq:H2} on medium-size lattices ($L=16$ and $L=20$),
performing full diagonalisation of the Hamiltonian, for two different
temperatures $1/\beta_{\rm Ising}$ of the Ising system in the ordered
phase, and for two choices of the spin-fermion coupling $\Lambda$. 

\begin{figure}[t]
  \centering
  \includegraphics[width=0.48\textwidth]{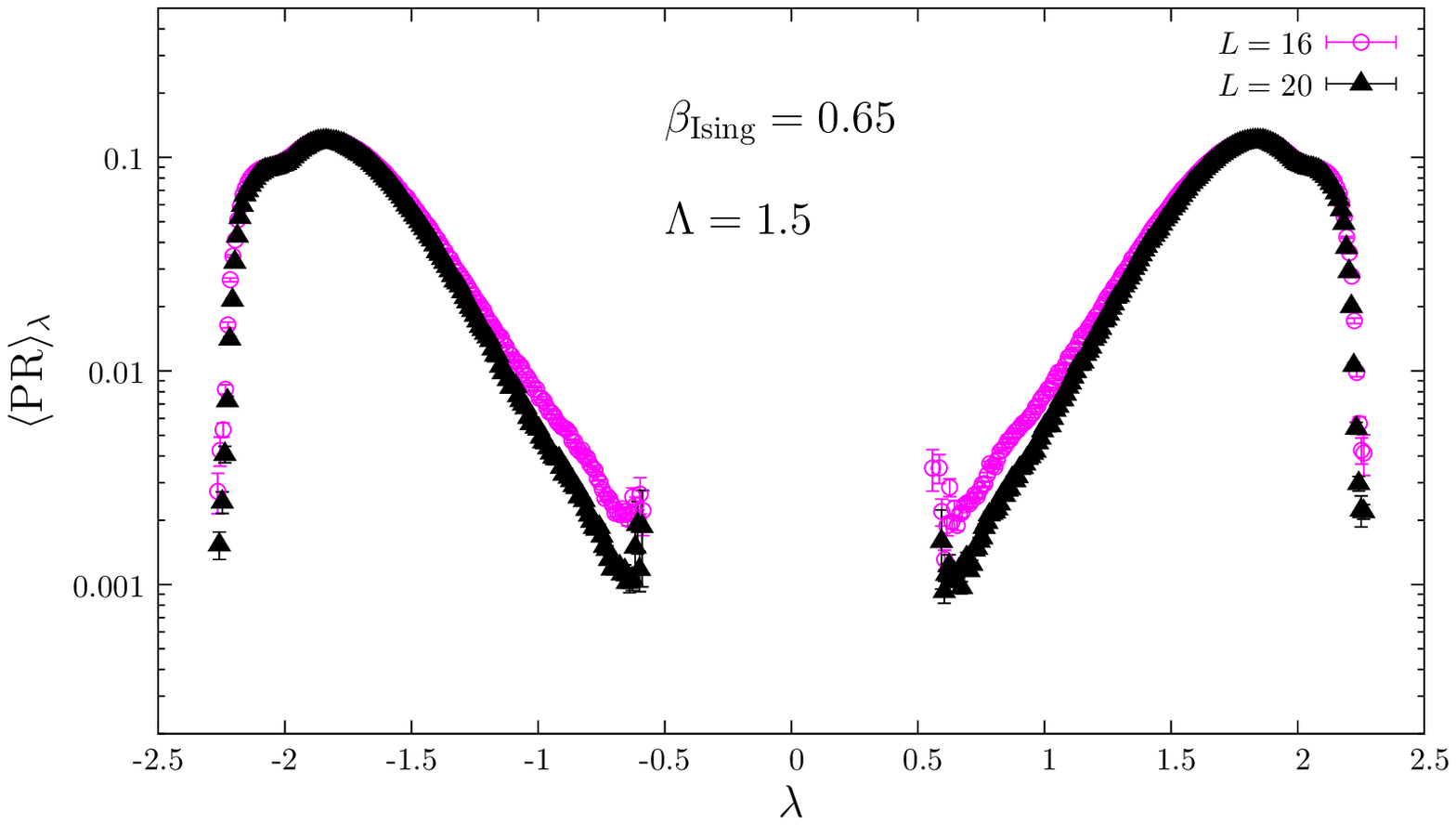}\hfil
  \includegraphics[width=0.48\textwidth]{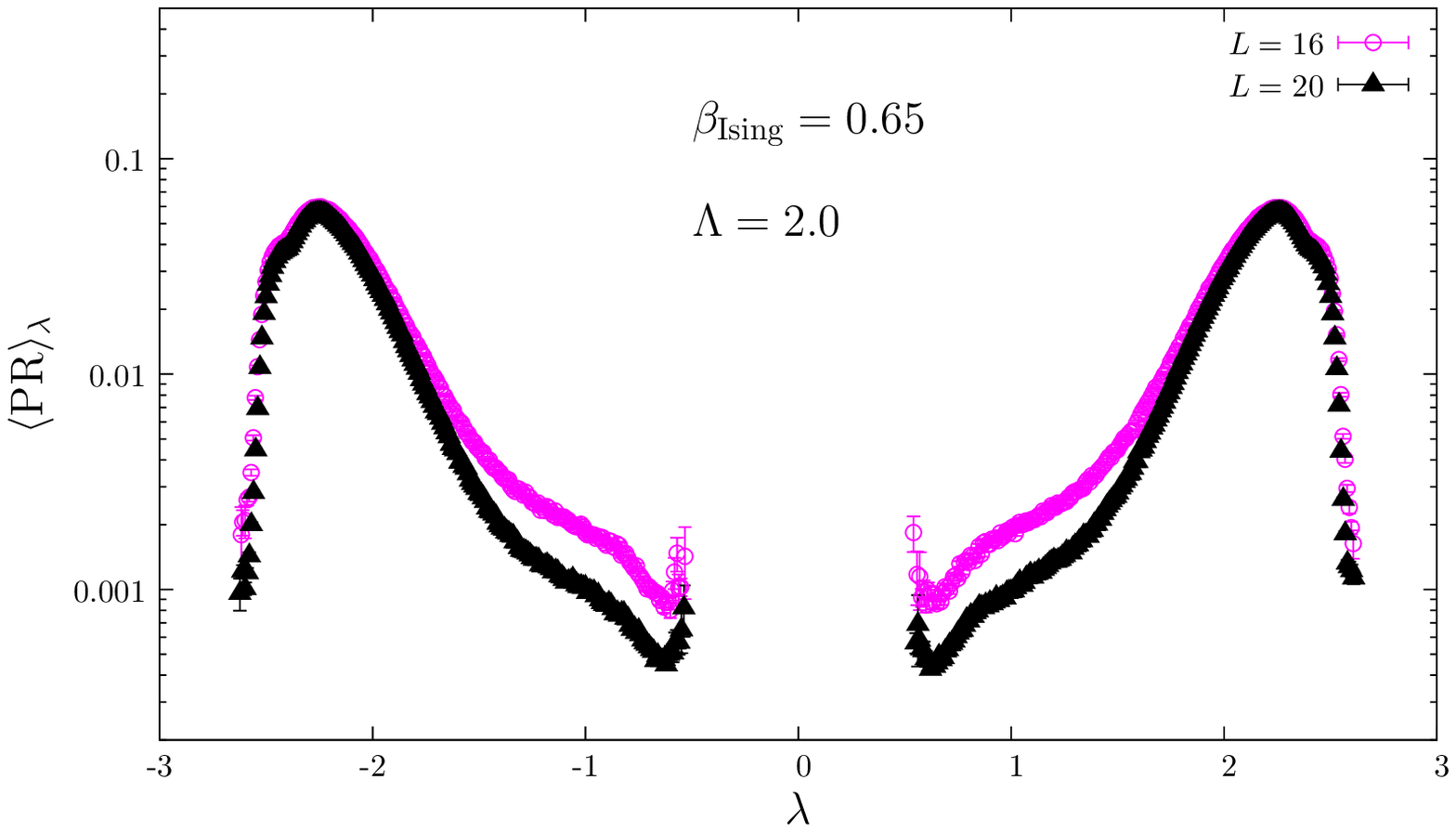}
  \includegraphics[width=0.48\textwidth]{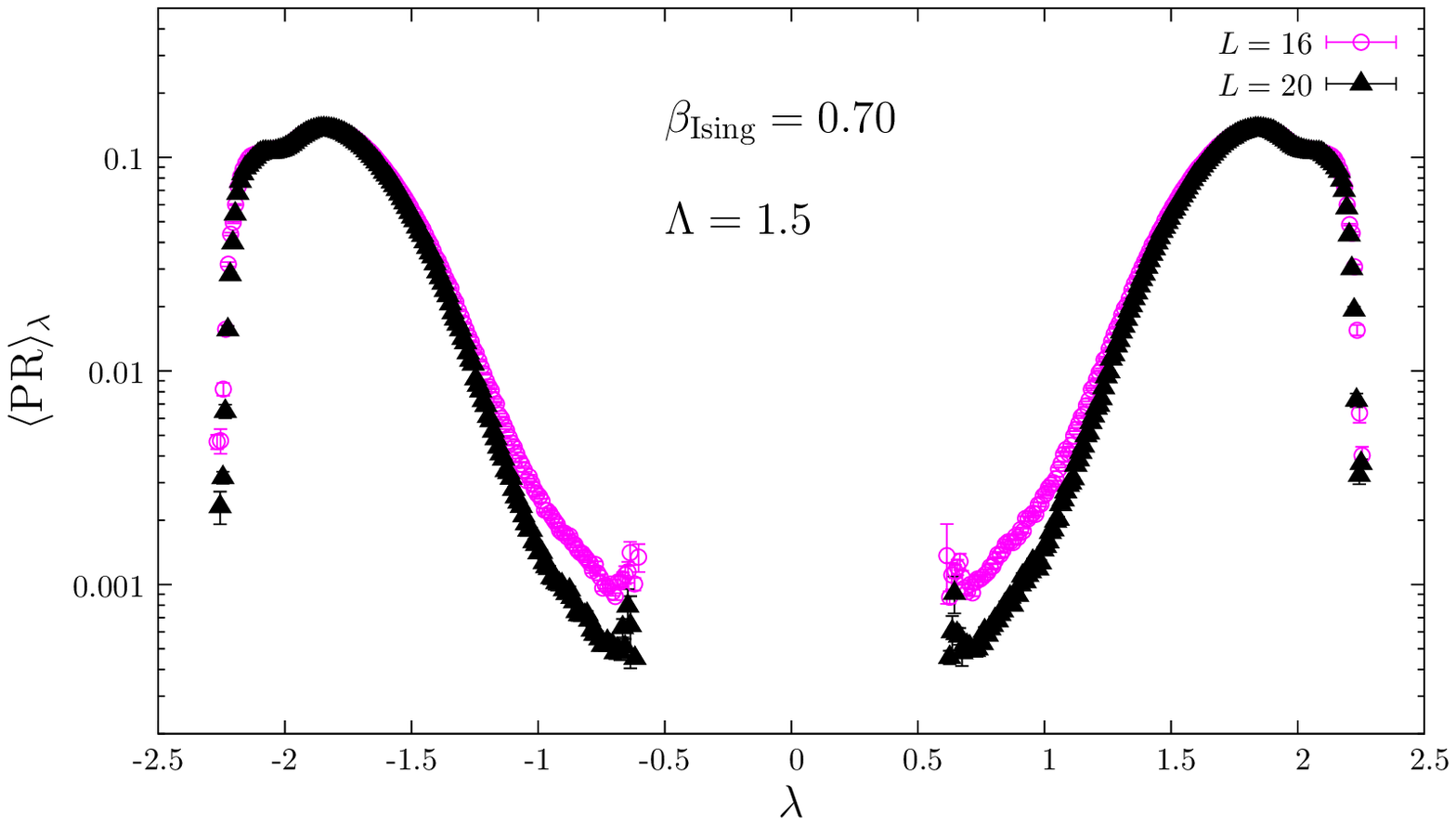}\hfil
  \includegraphics[width=0.48\textwidth]{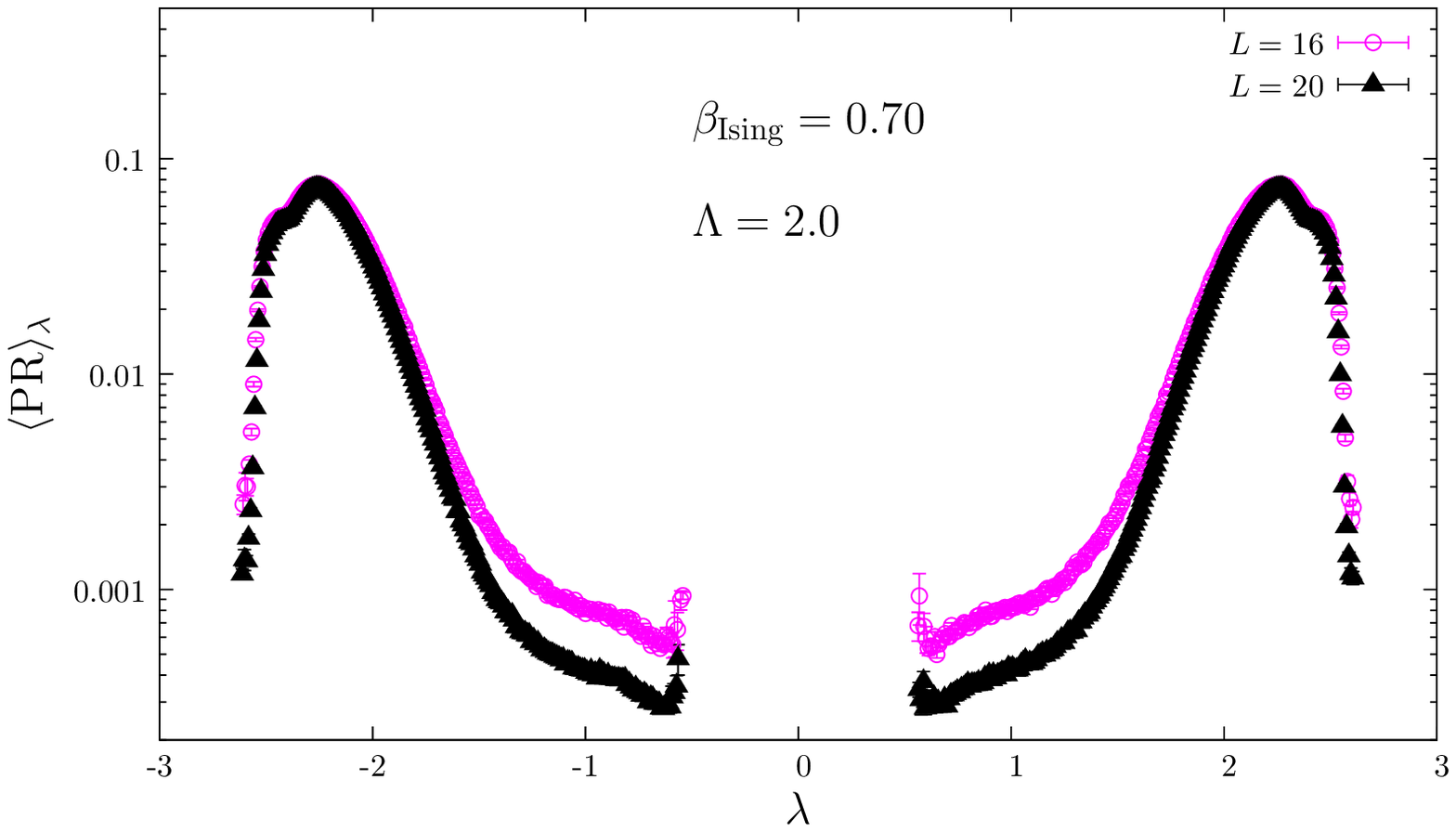}
  \caption{Participation ratio of eigenmodes of the effective model
    Eq.~\protect\eqref{eq:H2}.} 
  \label{fig:2}
\end{figure}

In Fig.~\ref{fig:1} we show the spectral density per unit volume of
the system. Low modes have small spectral density, which rapidly
increases as one goes up in the spectrum, as expected. Decreasing the
temperature, thus making the system more ordered, decreases the
density of low modes, as one expects if these modes are
localised. Increasing the spin-fermion coupling enlargens the region
where the spectral density is small, again as expected, since it
should push the effective gap up in the spectrum. Notice the symmetry
under $\lambda \to -\lambda$ of the spectrum, which can be proved to
be a property of the average spectrum, but which does not hold
configuration by configuration as it does in QCD with staggered
fermions. There is apparently also a sharp gap in the spectrum, which
is however essentially insensitive to $\Lambda$. It is not clear 
if this is due to the use of not large enough volumes or if it is
indeed a feature of the effective model, possibly due to the absence
of disorder in the hopping terms; in any case, it is irrelevant for
our purposes.

In Fig.~\ref{fig:2} we show the average participation ratio $PR=(\sum_x
|\psi(x)|^4)^{-1}/ V$, which gives a measure of the fraction of space
occupied by a given mode, as a function of its location in the
spectrum. The $PR$ changes by two orders of magnitude as one moves up
in the spectrum starting from the low-density region; more importantly, 
when increasing the size of the system it remains almost constant in
the bulk of the spectrum, while it visibly 
decreases near the origin. This signals that modes in the bulk are
delocalised, while modes near the origin are localised.

\begin{figure}[t]
  \centering
  \includegraphics[width=0.48\textwidth]{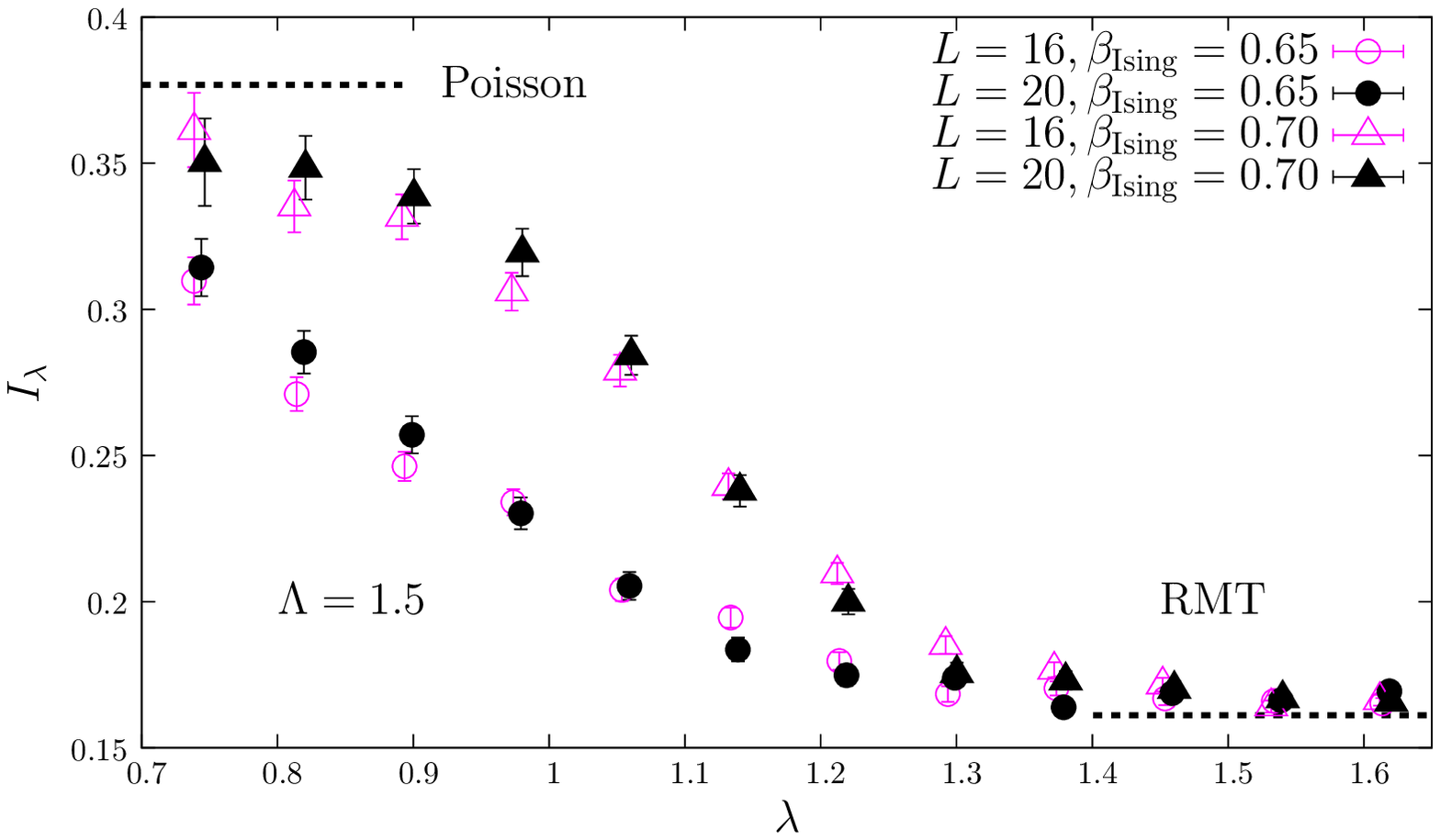}\hfil
  \includegraphics[width=0.48\textwidth]{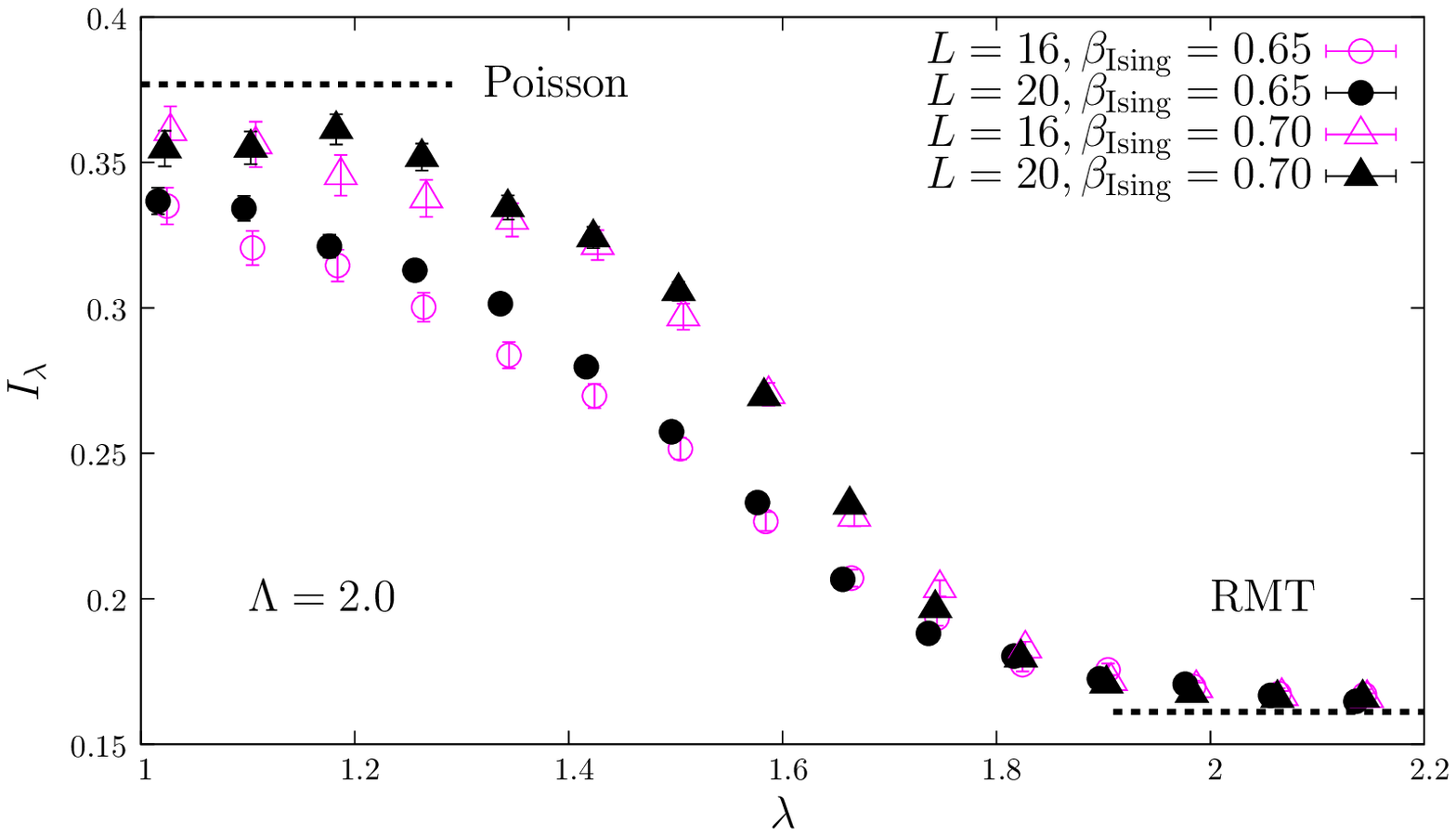}  
  \caption{The spectral statistics $I_\lambda$,
    Eq.~\protect\eqref{eq:ilam}, in the effective model 
    Eq.~\protect\eqref{eq:H2}.} 
  \label{fig:3}
\end{figure}

Finally, in Fig.~\ref{fig:3} we show a suitably defined local spectral
statistics $I_\lambda$ across the spectrum. Spectral
statistics can be used to detect a localisation/delocalisation
transition, since the eigenvalues corresponding to localised or
delocalised eigenmodes are expected to obey different statistics,
namely Poisson or Wigner-Dyson statistics, respectively. More
precisely, $I_\lambda$ is defined as  
\begin{equation}
  \label{eq:ilam}
I_\lambda = \int_0^{\bar s} ds\, p_\lambda(s)\,, \qquad
s_i=\f{\lambda_{i+1}-\lambda_i}{\la\lambda_{i+1}-\lambda_i\ra_\lambda}\,,\qquad
\bar s\simeq 0.5\,,  
\end{equation}
where $p_\lambda(s)$ is the probability distribution, computed locally
in the spectrum, of the so-called unfolded level spacing $s_i$, i.e., the 
level spacing divided by the local average level spacing
$\la\lambda_{i+1}-\lambda_i\ra_\lambda$. The results confirm 
that the eigenmodes change from localised to delocalised when
one moves up in the spectrum, and that the ``mobility edge''
separating localised and delocalised modes goes up in the spectrum
when the Ising system is made more ordered by decreasing the
temperature, or when the spin-fermion coupling is increased.
Furthermore, they give also a first indication that the steepness of
the curve increases as the volume is increased, thus hinting at the
existence of a true phase transition in the spectrum.

\section{Conclusions}

We have proposed a possible mechanism to explain localisation of quark
eigenmodes in QCD above $T_c$. The mechanism is based on the
``trapping'' effect of spatial fluctuations of the local Polyakov line
on the eigenmodes in the background of a partially ordered gauge
configuration. To test the explanation, we have constructed an
effective 3D model which should capture the main features of
localisation. Preliminary numerical results support the viability of
our explanation.

\end{document}